    \documentclass[pra,10pt,twocolumn]{revtex4}

\usepackage{amsmath} 
\usepackage{graphicx}
\usepackage{hyperref}
\hypersetup{colorlinks=true, linkcolor=blue, citecolor=blue, urlcolor=blue}

\begin{document}

 \def\BE{\begin{equation}}
 \def\EE{\end{equation}}
 \def\BA{\begin{array}}
 \def\EA{\end{array}}
 \def\BEA{\begin{eqnarray}}
 \def\EEA{\end{eqnarray}}
 \def\nn{\nonumber}
 \def\ra{\rangle}
 \def\la{\langle}

 \title{Quantum statistics of Schr{\"o}dinger cat states prepared by logical gate with non-Gaussian resource state}
 \author{N.~I.~Masalaeva}
 \author{I.~V.~Sokolov}
 \email{i.sokolov@spbu.ru, sokolov.i.v@gmail.com}
 \affiliation{Saint Petersburg State University, Universitetskaya nab.  7/9, 199034  Saint Petersburg, Russia}


\begin{abstract}
A measurement-induced continuous-variable logical gate is able to prepare Schr{\"o}dinger cat states if the gate uses a non-Gaussian resource state, such as  cubic phase state [I. V. Sokolov, Phys. Lett. A {\bf 384}, 126762 (2020)].
Our scheme provides an alternative to hybrid circuits which use photon subtraction and (or) Fock resource states
and photon number detectors.
We reveal the conditions under which the gate conditionally prepares quantum superposition of two undistorted
``copies'' of an arbitrary input state that occupies a finite area in phase space. A detailed analysis of the fidelity
between the gate output state and high-quality Schr{\"o}dinger cat state is performed. A clear interpretation
of the output state quantum statistics in terms of Wigner function in dependence on the gate parameters and
measurement outcome is presented for a representative set of input Fock states.
\end{abstract}


 \maketitle

 \section{Introduction}

The continuous-variable (CV) quantum information schemes based on Gaussian resource states were extensively explored both theoretically and experimentally~\cite{Lloyd99,Braunstein05,Furusawa98,Li02}, including their essentially multimode implementation~\cite{Gu09,Weedbrook12,Ukai15,Yokoyama13,Roslund14}. While the continuous-variable Gaussian cluster schemes are able to perform Gaussian transformations of the input states, in order to achieve universal quantum computing, there is a need to introduce the non-Gaussian logical gates ~\cite{Lloyd99,Braunstein05}.

A minimal nonlinearity sufficient to prepare non-Gaussian resource states is cubic. The cubic phase state based on the cubic nonlinearity was first considered in~\cite{Gottesman01,Bartlett_Sanders02}. Various approaches to the implementation of such states, as well as of the cubic (or higher) phase gates, were explored both theoretically and experimentally~\cite{Ghose07,Marek11,Yukawa13,Marshall15,Miyata16,Marek18}.

It was shown recently~\cite{Sokolov20}, that continuous-variable measurement-induced two-node logical gate is able to prepare Schr{\"o}dinger cat-like quantum superpositions, if Gaussian (e. g. squeezed) resource state of an ancillary oscillator is substituted with non-Gaussian one. This was demonstrated for the cubic phase state used as a resource.

An important distinction between the gate~\cite{Sokolov20} and the non-Gaussian circuits mentioned above is that the Schr{\"o}dinger cat state emerges when the ancillary oscillator measurement is compatible not with one, but with two different values of the target oscillator physical variables. This feature does not appear in Gaussian quantum networks and, given that a non-Gaussian resource state is used, needs the measurement which is consistent with this criterion.

In general, one can prepare CV Schr{\"o}dinger cat states by making use of the unitary evolution assisted by a non-linear interaction~\cite{Yurke86}. The schemes based on a hybrid measurement-induced evolution also can create cat-like states.
The optical Schr{\"o}dinger cat states were generated in a low-photon regime using photon subtraction
\cite{Ourjoumtsev06}, the homodyne detection with photon number state as a resource~\cite{Ourjoumtsev07}, and
the iterative schemes which allow an incremental enlargement of cat states \cite{Etesse14,Etesse15,Sychev17}.
The latter proposals are specifically aimed at the creation of superpositions of two coherent states. The
Schr{\"o}dinger cat states can be prepared by using a photonic even-parity detector and CV entanglement
\cite{Thekaddath20}.

The Schr{\"o}dinger cat states are an object of relentless interest since their first introduction~\cite{Schrodinger35}. Besides their fundamental importance, some proposals for fault-tolerant quantum information processing directly rely on the cat-like states~\cite{Gottesman01,Mirrahimi14}.

In this work, we explore in detail the quantum statistics of the superpositions, which are prepared by our gate from an arbitrary input state that occupies a limited area in phase space. Of interest is the gate operation regime, where the gate output state is close to  ``perfect'' Schr{\"o}dinger cat state, that is, to the superposition of two symmetrically displaced undistorted copies of the input state. By considering the fidelity between the actual gate output state and the needed form of cat state, we find the conditions under which the gate prepares high-quality cat states.

We present the output states Wigner functions for a wide range of the gate parameters and reveal the connection between the fidelity and the shape of the input state copies support in phase space.

While the gate is universal and can prepare Schr{\"o}dinger cat states from almost arbitrary input states, the quality of the output states depends on the input states extent along the coordinate and momentum axes. We consider input Fock states with up to two photons and demonstrate the gate potential to prepare high-quality cat-like superpositions out of these and similar states.

Note that considerable efforts have been made to generate superpositions of coherent states with large enough distance between the components~\cite{Ourjoumtsev06,Ourjoumtsev07,Etesse14,Etesse15,Sychev17}. The gate based on the cubic phase state can prepare such superpositions with as large as needed distance between the copies in phase space.

\par

 \begin{figure*}
 \begin{center}
 \includegraphics[width=0.95\textwidth]{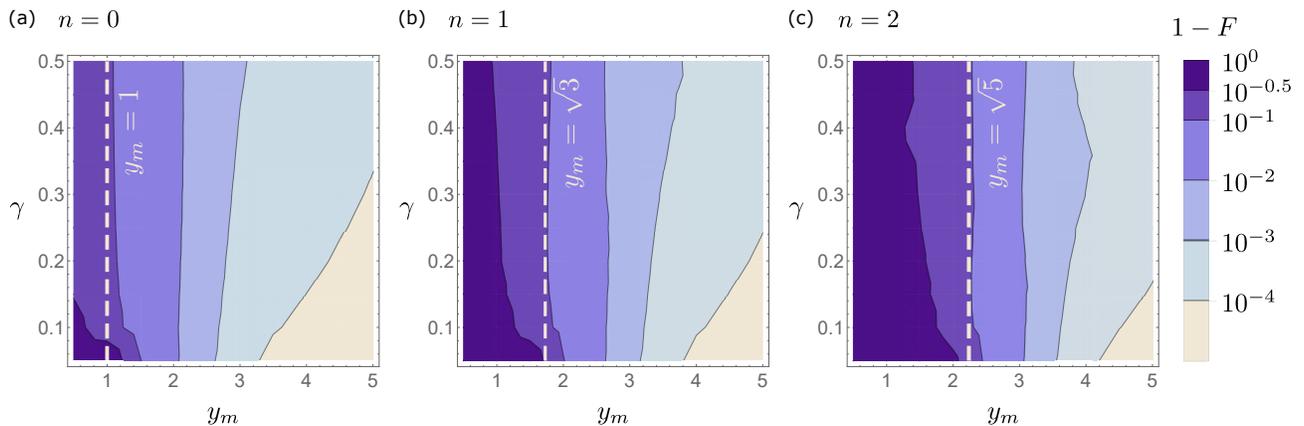}
 \caption{Infidelity between the exact solution~\eqref{out} for the output state and the output state~\eqref{cat_stphase} evaluated in the stationary phase approach. The input states are Fock states with $n=0, 1, 2$ quanta correspondingly. The vertical dashed lines $y_m=\sqrt{2n+1}$ separate the areas where the stationary phase approach does not work.}
 \label{fig_St_fidelity}
 \end{center}
 \end{figure*}

\par

Our analysis is based on the exact form of the output state~\cite{Sokolov20}. At the same time, the gate operation may be illustrated in terms of an approximate visual representation of the step-by-step quadrature amplitudes transformations in the scheme if one considers these transformations in a semiclassical manner. This approach gives a clear idea how Schr{\"o}dinger cat states may emerge in similar circuits. It might be useful for the analysis of logical gates based on more complex non-Gaussian resource states, where exact solution is not available.


\section{Cat-like states from CV gate with non-Gaussian resource state}

The non-Gaussian gate we discuss here was introduced in~\cite{Sokolov20}. The measurement-induced two-node gate uses the cubic phase state as an elementary non-Gaussian resource, the entangling $C_Z$ operation, and the projecting homodyne measurement.

A key feature of the gate is that the measurement outcome provides multivalued information about the output state canonical variables, which results in the preparation of a cat-like state. This feature is interpreted in the next section in terms of a clear pictorial representation.

Let us outline briefly the gate operation (for more details see~\cite{Sokolov20}). The target oscillator is initially prepared in an arbitrary state
 $$
|\psi_1\ra = \int dx_1\psi(x_1)|x_1\ra,
 $$
which is assumed to occupy a limited range $\{\Delta x_1,\,\Delta y_1\}$ of the coordinate and momentum. In order to prepare the non-Gaussian resource state of the ancillary oscillator, one applies unitary non-Gaussian evolution operator $\exp(i\gamma q_2^3)$ to the momentum eigenstate $|0\ra_{p_2}$,
 $$
|\psi_2\ra = e^{i\gamma q_2^3}|0\ra_{p_2} = \int dx_2e^{i\gamma x_2^3}|x_2\ra.
 $$
In the following, we use the notation $\{q,\,p\}$  for the coordinate and momentum operators.

The $C_Z$ entangling unitary evolution operator $\exp(iq_1q_2)$ prepares the state
 \BE
 \label{entangled}
 |\psi_{12}\ra = \int dx_1dx_2\psi(x_1)e^{ix_2(x_1 + \gamma x_2^2)}|x_1\ra|x_2\ra.
 \EE
Next, the ancillary oscillator momentum is measured with the outcome $y_m$. Projecting the state~\eqref{entangled} on the homodyne detector eigenstate $|y_m\ra_{p_2}$, one arrives at target oscillator output state
 \BE
 \label{out}
\psi^{(out)}(x) = \sqrt{\cal N} \psi(x)\varphi(x - y_m),
 \EE
where $x_1 \to x$  for brevity, and ${\cal N}$ is the normalization factor. As the result of gate operation, the input state is multiplied by the factor
\begin{multline}
\varphi(x-y_m) = \frac{1}{\sqrt{2\pi}}\int dx'e^{ix'(x - y_m + \gamma x'^2)}\\
=\big[\sqrt{2\pi}/(3\gamma)^{1/3}\big] {\rm Ai}\big[(x-y_m)/(3\gamma)^{1/3}\big],\label{exact}
\end{multline}
which is expressed in terms of the Airy function.

While the result~\eqref{out},~\eqref{exact} is exact, it is instructive to consider the approximation where one can specify two stationary phase points of the exponent in~\eqref{exact}. In this approximation, the added factor becomes a sum of two contributions which directly correspond to the components of the output quantum superposition. In the following, we demonstrate that cat-like states arise just if this approximation is valid.

For a given target oscillator coordinate $x$, the stationary points are
 \BE
 \label{stphase_points}
x'_{st}=\pm\sqrt{(y_m-x)/3\gamma}.
 \EE
If two stationary phase points exist for all $x$ within the region spanned by the input state, we finally arrive at the approximate output state
 \BE
 \label{cat_stphase}
 \psi^{(out)}_{st}(x) = \sqrt{{\cal N}_{st}}\psi(x)\big[\varphi^{(+)}(x - y_m) + c.\, c.\big],
 \EE
where ${\cal N}_{st}$ is the normalization factor, and
\begin{multline}
\varphi^{(+)}(x - y_m) = \exp \Big\{ i\big[\frac{\pi}{4} - \frac{2}{3\sqrt{3\gamma}}(y_m - x)^{3/2}\big]\Big\}\\
\times\big[12\gamma(y_m - x)\big]^{-1/4}.\label{factor_stphase}
\end{multline}

In Fig.~\ref{fig_St_fidelity} we plot the infidelity $1 - F_{st}$ between the exact solution~\eqref{out} and the output state~\eqref{cat_stphase} evaluated in the stationary phase approach, where the fidelity is
 \BE
 \label{st_fidelity}
F_{st} = \left| \int dx \psi^{(out)*}(x)\psi^{(out)}_{st}(x)\right|^2.
 \EE
As the input states of the target oscillator, we choose here the first Fock states $|n\ra$ with the coordinate wave function
 \BE
 \label{Fock}
\psi^F_n(x) = \frac{1}{\pi^{1/4}\sqrt{2^n n!}} H_n(x)e^{-x^2/2},
 \EE
where $H_n(x)$ it the Hermite polynomial, and $n=0, 1, 2$ correspondingly.

If the input state probability distribution is concentrated within a certain range $-\Delta x \leq x \leq \Delta x$,
and the measured ancilla momentum is not large enough, $y_m \leq \Delta x$, there are coordinates $x$ such
that two distinct stationary phase points do not exist, see~\eqref{stphase_points}. For the vacuum input state one
can assume $\Delta x \to 1$. The fidelity~\eqref{st_fidelity} for $y_m \leq 1$ is low, as seen from the Fig.~\ref{fig_St_fidelity}(a). At the same time, beyond the region $y_m \leq 1$ the stationary phase approach
provides higher fidelities.

For the input Fock states $|n\ra$ with $n$ = 1 and 2, better fidelities are achieved in the area $y_m \geq \sqrt{2n+1}$, see Fig.~\ref{fig_St_fidelity}(b, c), in agreement with the estimate $\Delta x \sim \sqrt{2n+1}$.


 \section{Input state evolution in phase space}

\begin{figure}
 \begin{center}
 \includegraphics[width=0.8\columnwidth]{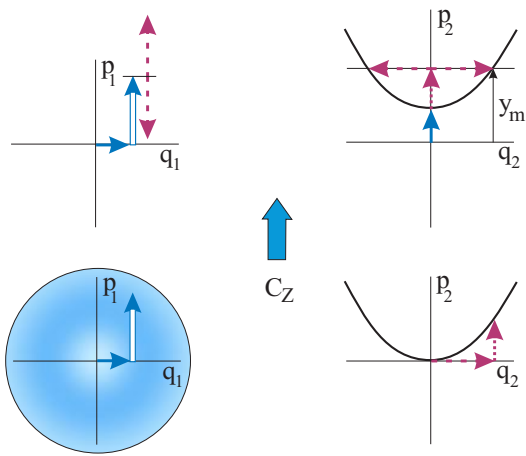}
 \caption{Measurement-induced evolution of quadrature amplitudes of the target (left) and ancillary (right) oscillators. The randomly chosen initial amplitudes undergo the transformation of the cubic phase state preparation (bottom right), the entangling non-demolition $C_Z$ operation, and the ancilla momentum measurement with the outcome $y_m$ (top right). The target oscillator state collapses to the cat-like state (top left).}
 \label{fig_Scheme}
 \end{center}
 \end{figure}

There is a clear pictorial representation~\cite{Sokolov20} in phase space of the evolution performed by the gate, illustrated in Fig.~\ref{fig_Scheme}. The scheme shows the step-by-step transformation of the target and ancillary oscillators quadrature amplitudes $\{q_i,\,p_i\}$, where $i=1,2$, as these transformations arise in the Heisenberg picture under some simplifying admissions.

The initial canonical variables of both nodes of our scheme are denoted via $\{q_i(0),\,p_i(0)\}$. The ancilla cubic phase state is prepared by the non-Gaussian unitary evolution operator $\exp(i\gamma q_2^3)$ applied to the initial ancilla momentum eigenstate $|0\ra_{p_2}$. The Heisenberg equations for canonical variables solved on unit time interval with the Hamiltonian $-\gamma q_2^3$ yield,
 \BE
 \label{p_of_q_semicl}
 \BA{l} q_1 = q_1(0), \\p_1 = p_1(0),\EA \qquad
 \BA{l} q_2 = q_2(0), \\p_2 = p_2(0) + 3\gamma q_2^2(0).\EA
 \EE
Next, the two-node entangling evolution $\exp(iq_1q_2)$ (the non-demolition $C_Z$ operation) is applied, which adds coordinate of the 1st oscillator to the momentum of the 2nd, and vice versa. The canonical variables in the entangled state are found to be
 \BE
 \label{qp_inout}
\BA{ll}q^e_1=q_1(0),\\p^e_1=p_1(0)+q_2(0),\\ \\
q^e_2=q_2(0),\\p^e_2 = p_2(0) + 3\gamma q_2^2(0)+q_1(0).\EA
 \EE
In Fig.~\ref{fig_Scheme}, we represent these transformations by randomly choosing initial points in phase space
for both oscillators within the region
spanned by the corresponding statistical ensemble. The initial ancilla state $|0\ra_{p_2}$ with zero momentum uncertainty may be represented by a horizontal line. The transformation~\eqref{p_of_q_semicl} prepares the
well-known non-Gaussian cubic phase state~\cite{Gottesman01, Bartlett_Sanders02}
(up to the $q \leftrightarrow p$ substitute), whose Wigner function demonstrates fringes and
negativity~\cite{Ghose07}, and the Wigner logarithmic negativity is almost maximal among some elementary
non-Gaussian states~\cite{Albarelli18}.

Nevertheless, the gate under consideration can be effectively described if the transformations~\eqref{p_of_q_semicl},~\eqref{qp_inout} and~\eqref{cubic_cat_transf} (see below) are treated in a semiclassical manner, that is, when it is admitted that one can put in correspondence a certain value of momentum
to any given coordinate. In this approximation, the horizontal line which represents the ancilla initial state is mapped point-by-point by the transformation~\eqref{p_of_q_semicl} to the parabola shown in Fig.~\ref{fig_Scheme} (bottom right).

The ancilla momentum measurement with the outcome $y_m$ imposes links to the possible values of the coordinates
and momenta, as shown in Fig.~\ref{fig_Scheme} (top right). By substituting the observed momentum value
$y_m$ for $p^e_2$, one arrives at the following measurement-induced target oscillator quadrature amplitudes,
depicted in the same figure (top left),
 \BE
 \label{cubic_cat_transf}
 \BA{ll}q^{(out)}_1=q_1(0),\\p^{(out)}_1 = p_1(0) \pm (3\gamma)^{-1/2}\sqrt{y_m - q_1(0) - p_2(0)}.\EA
 \EE
In view of $p_2(0)|0\ra_{p_2} = 0$, one can drop $p_2(0)$ here.

In general, substituting the measurement outcome for the relevant operator-valued observable in the Heisenberg
picture may ruin the commutation relations. As far as the output variables are multivalued in our scheme, the result~\eqref{cubic_cat_transf} can not be considered as exact in the Heisenberg picture.

 \begin{figure}
 \begin{center}
 \includegraphics[width=0.95\columnwidth]{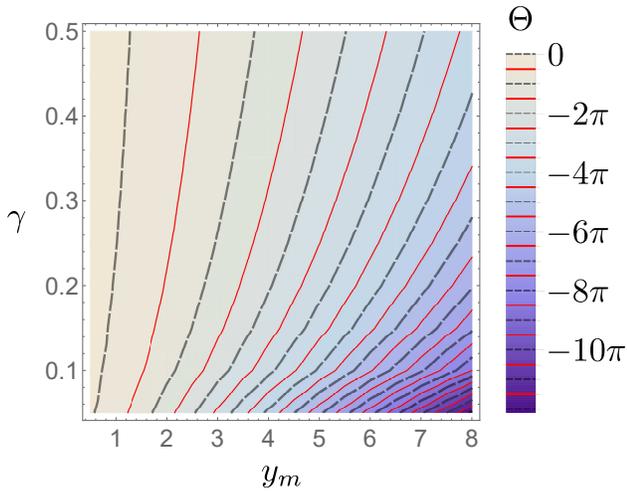}
 \caption{Phase $\theta$~\eqref{theta} as a function of the gate parameters $\gamma$
 and measurement outcome $y_m$. The black dashed and red solid lines correspond to even and odd cat states, respectively.}
 \label{fig_Theta}
 \end{center}
 \end{figure}

Nevertheless, the output amplitudes~\eqref{cubic_cat_transf} agree not only with the semiclassical visualization
in phase space of the transformation performed by the gate but also with the solution~\eqref{cat_stphase}, as
we demonstrate below. Note that two distinct values of the output oscillator momentum arise just if there are
two distinct stationary phase points. The distance between two values of $x'_{st}$ (which are eventually mapped
to the output momentum due to the entanglement), and between two values of $p^{(out)}_1$
in~\eqref{cubic_cat_transf} is given by the same expression for any output coordinate.

That is, there is a direct correspondence between the Heisenberg-like picture we discuss here and its
visualization on the one hand, and the solution derived in the stationary phase approximation on the other hand.

In general, a cat-like state is prepared when the ancilla measurement outcome is compatible with two or more
distinct values of the target oscillator variables. This feature does not arise in the Gaussian measurement-induced
CV schemes since it is due to the non-linear deformation of the resource state in phase space. With the cubic
phase state used as a non-Gaussian resource,  two ``copies'' of an input state may be created by the gate, as seen from~\eqref{factor_stphase} and~\eqref{cubic_cat_transf}.


\section{Quantum statistical properties of the output Schr{\"o}dinger cat states}

In this section, we discuss the potential of our scheme to prepare ``perfect''  cat states, where two components of
the output superposition are undistorted copies of the input state symmetrically displaced in phase space along
the momentum axis.

For the input state whose coordinate wave function is concentrated around  $x=0$, one can Taylor expand the
added factor~\eqref{factor_stphase} in $x$. In the linear in $x$ approximation for the exponent, we arrive at
 \BE
 \label{phi_cat}
\varphi^{(+)} \to \varphi^{(+)}_{cat} = \exp[i(\theta + p^{(+)} x)](12y_m\gamma)^{-1/4},
 \EE
where
 \BE
 \label{theta}
\theta =  \frac{\pi}{4} - \frac{2}{3\sqrt{3\gamma}}y_m^{3/2}, \qquad p^{(+)} = \sqrt{y_m/3\gamma}.
 \EE
Here  $p^{(+)}$ is the displacement of a copy due to the factor $\varphi^{(+)}$.

 \begin{figure}
 \begin{center}
 \includegraphics[width=0.7\columnwidth]{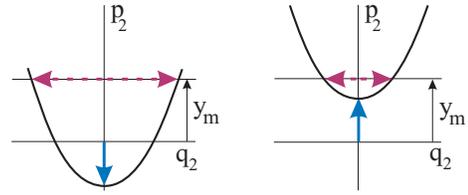}
 \caption{The measurement-induced splitting of the observed ancilla coordinate  (dashed arrows) for different
 values of the target oscillator coordinate $q_1$ (solid arrows).}
 \label{fig_Intersections}
 \end{center}
 \end{figure}
 \begin{figure*}
 \begin{center}
 \includegraphics[width=0.95\textwidth]{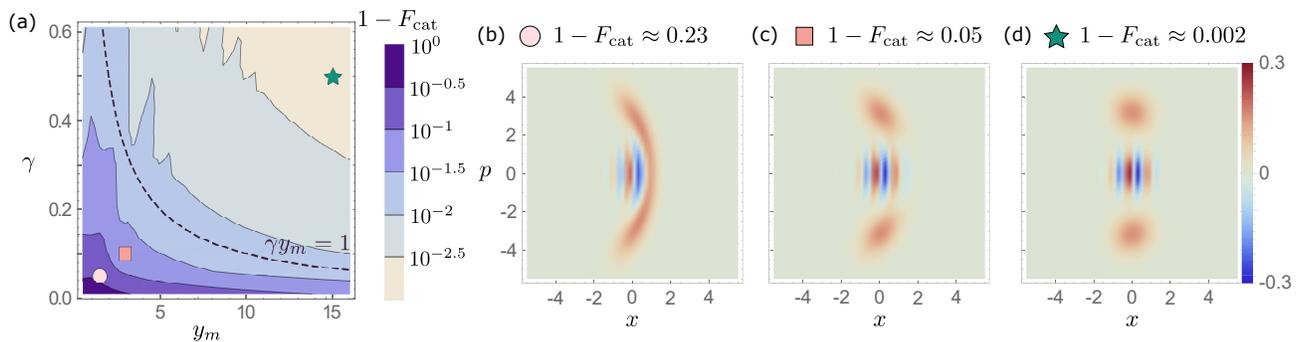}
 \caption{Infidelity 1 - $F_{cat}$ (a) between the exact output state~\eqref{out} and ``perfect'' cat state~\eqref{cat_coord} for vacuum input state of the target oscillator as a function of the gate parameter
 $\gamma$ and measurement outcome $y_m$. The  Wigner function of the exact output state  for the points
 $\{y_m,\gamma\}$ indicated by the circle (b), square (c), and asterisk (d), respectively.}
 \label{fig_n0}
 \end{center}
 \end{figure*}
Consider cat state preparation from the input Fock state~\eqref{Fock}. In the approximation adopted here,
the output state coordinate wave function is given by
 \BE
 \label{cat_coord}
\psi^{(out)}_{n,cat}(x) =
\sqrt{{\cal N}_{n,cat}}\cos\big(\theta + \sqrt{y_m/3\gamma}\, x\big) H_n(x)e^{-x^2/2},
 \EE
where ${\cal N}_{n,cat}$ is the normalization factor. In particular, the displaced vacuum input state may be
represented in terms of the Glauber state $|\alpha\ra$ as
 $$
\exp[i(\theta + \sqrt{y_m/3\gamma} x)]\psi^F_0(x) \leftrightarrow e^{i\theta} |\alpha\ra,
 $$
where $\alpha = i\sqrt{y_m/6\gamma}$. Hence, the Schr{\"o}dinger cat state produced from the vacuum
one is
 \BE
 \label{cat_Glauber}
\sqrt{{\cal N}_{vac}}\left(e^{i\theta} |\alpha\ra + e^{-i\theta}|-\alpha\ra\right),
 \EE
where
 $$
{\cal N}^{-1}_{vac} = 2\big[1 + \cos(2\theta) e^{-2|\alpha|^2}\big].
 $$
In the coordinate representation, the state~\eqref{cat_Glauber} is
 \BE
 \label{cat_Glauber_coord}
\psi^{(out)}_{0,cat}(x) = \frac{\sqrt{2}}{\pi^{1/4}}
\frac{\cos\big(\theta + \sqrt{y_m/3\gamma}\, x\big)}{\sqrt{(1 + \cos(2\theta) e^{-y_m/3\gamma})}} e^{-x^2/2}.
 \EE
Given that the approximation~\eqref{phi_cat} is valid, an arbitrary input state is transformed by the gate to
the cat-like superposition of two undistorted copies symmetrically shifted along the momentum axis, in analogy to~\eqref{cat_coord} and~\eqref{cat_Glauber}. The phase $\theta$~\eqref{theta} which determines the relative
phase shift of the cat state components  is depicted in Fig.~\ref{fig_Theta} as a function of the gate parameters
and measurement outcome. The emerging cat state is even for $\theta=m \pi$ and odd for
$\theta=(m+1/2)\pi$ with m integer.

It is instructive to evaluate  the deformation of the cat state components and to reveal optimal conditions under
which the gate prepares ``perfect'' Schr{\"o}dinger cat states. In Fig.~\ref{fig_Intersections}, we illustrate
how such deformation arises for the cubic phase state used as a non-Gaussian resource state in our scheme.
This figure reproduces the top right part of Fig.~\ref{fig_Scheme} for different target oscillator coordinates.
As seen from Fig.~\ref{fig_Intersections}, for a given ancilla momentum measurement outcome, the
arising splitting of the observed ancilla coordinate may essentially depend on the target oscillator coordinate.

 \begin{figure*}
 \begin{center}
 \includegraphics[width=0.95\textwidth]{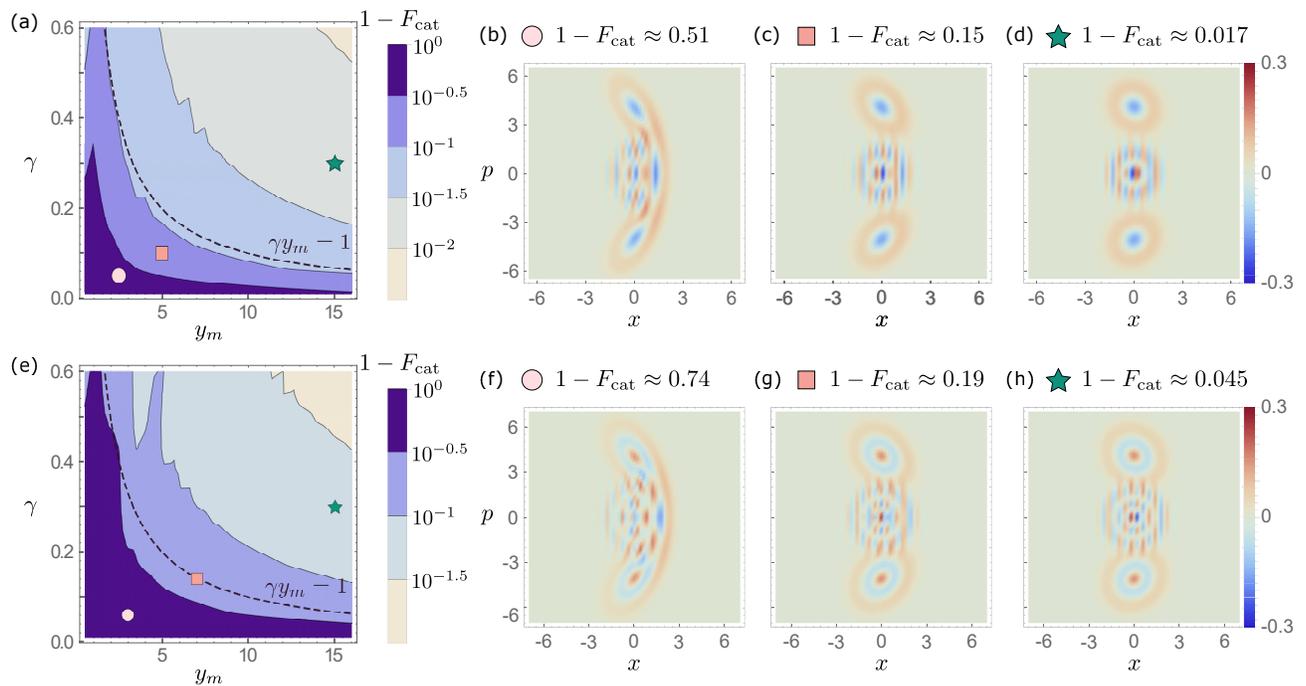}
 \caption{Infidelity 1 - $F_{cat}$ between the exact output state~\eqref{exact} and ``perfect'' cat
 state~\eqref{cat_coord} for the single-photon (a) and two-photon (e) input state as a function of the gate
 parameter $\gamma$ and measurement outcome $y_m$. The Wigner function of the exact output state for the single-photon (b,c,d) and two-photon (f,g,h) input states for the points $\{y_m,\gamma\}$ indicated by the
 circle, square, and asterisk, respectively.}
 \label{fig_n1_n2}
 \end{center}
 \end{figure*}

In order to estimate the linear shearing deformation of the copies, we take into account the next term in
the Taylor expanded phase of the added factor~\eqref{factor_stphase},
 $$
\varphi^{(+)} \sim \exp\Big\{i\big[\theta + (p^{(+)} + \delta p^{(+)}(x))x\big]\Big\},
 $$
where
 $$
\delta p^{(+)}(x) \approx - \frac{1}{4\sqrt{3\gamma y_m}} x.
 $$
Consider an input state which spans a range $2\Delta x$  along the coordinate axis. The linear in $x$
contribution $|\delta p^{(+)}(x)|$ to the momentum of a copy changes at the distance $2\Delta x$ as
 $$
|\Delta(\delta p^{(+)}(x))| \sim \lambda\cdot 2\Delta x,
 $$
where the parameter
 \BE
 \label{lambda}
\lambda = \frac{1}{4\sqrt{3\gamma y_m}} \sim \frac{0.14}{\sqrt{\gamma y_m}} ,
 \EE
can be viewed at as a measure of the linear shearing deformation.

In Figs.~\ref{fig_n0}(a) and~\ref{fig_n1_n2}(a,e) we represent the infidelity $1-F_{cat}$ between the exact
output state~\eqref{out} and ``perfect'' cat state~\eqref{cat_coord}, where
 \BE
 \label{cat_fidelity}
F_{cat} = \left| \int dx \psi^{(out)*}(x)\psi^{(out)}_{cat}(x)\right|^2,
 \EE
for input Fock states~\eqref{Fock} with $n$=0 and $n$ =1, 2, respectively. In these figures, we also plot the
curves $\gamma y_m=1$, thus separating the areas of ``large'' ($\lambda \geq 0.14$) and ``small''
($\lambda \leq 0.14$) deformation of the cat state components.

As seen from the figures, the fidelity $F_{cat}$ dependence on the gate parameters and measurement outcome
is in a good agreement with the estimate given above, and the gate output may be very close to
the ``perfect'' cat state.

In order to show how the output state looks like for the cases of bad, fair, and good fidelity~\eqref{cat_fidelity},
we plot in Figs.~\ref{fig_n0}(b,c,d),~\ref{fig_n1_n2}(b,c,d), and~\ref{fig_n1_n2}(f,g,h) the corresponding
output state Wigner functions evaluated using the exact solution~\eqref{exact}. The points in the plane
$\{y_m,\gamma\}$ indicated by a circle, square, and asterisk, respectively, are chosen in the regions of
a bad, fair, and good fidelity.

Consider first the vacuum input state, Fig.~\ref{fig_n0}(b,c,d). The points in the plane $\{y_m,\gamma\}$
for which the Wigner function was evaluated, are $\{1.5, 0.05\}$, $\{3, 0.1\}$, and $\{15, 0.5\}$. For these points,
the average shift $p^{(+)} = \sqrt{y_m/3\gamma}$ of a copy along the momentum axis is equal to $3.16$.
This shift is larger than the input state spread in phase space, and one may consider the output state as a
``real'' cat as opposed to ``kittens'' with small copies separation~\cite{Ourjoumtsev06}.

The Wigner function depicted in Fig.~\ref{fig_n0}(b) was evaluated for $\{y_m,\gamma\} = \{1.5, 0.05\}$
and $F_{cat}=0.77$. For this point, the fidelity $F_{st}$ between the exact solution and the solution
based on the stationary phase approach is low (see Fig.~\ref{fig_St_fidelity}(a)), and even within the latter
approach, the deformation parameter~\eqref{lambda} is large enough, $\lambda \approx 0.51$.
Fig.~\ref{fig_n0}(b) shows that the support of the output state Wigner function is split by the  gate,
but its shape hardly corresponds to ``perfect'' cat state.

The Wigner function shown in Fig.~\ref{fig_n0}(c) was evaluated for $\{y_m,\gamma\} = \{3, 0.1\}$
and $F_{cat}=0.946$. The plot demonstrates well separated but distorted copies of the input state. Finally, in Fig.~\ref{fig_n0}(d) we depict the output state Wigner function for $\{y_m,\gamma\} = \{15, 0.5\}$,
where high fidelity $F_{cat}=0.998$ is achieved, and the gate prepares the cat-like superposition close
to the ``perfect'' cat state~\eqref{cat_Glauber}.

If the input Fock state is single- or two-photon, the gate is also able to prepare good cat states as shown
in Figs.~\ref{fig_n1_n2}(b,c,d) and~\ref{fig_n1_n2}(f,g,h). Since $\Delta x \sim \sqrt{2n+1}$,
we choose for these states such parameters of the gate that ensure sufficient separation of the
copies.

For the single-photon input state (Figs.~\ref{fig_n1_n2}(b,c,d)), we take for $\{y_m,\gamma\}$ the values
$\{2.5, 0.05\}$, $\{5, 0.1\}$, and $\{15, 0.3\}$. The copy shift along the momentum axis is $p^{(+)} = 4.08$.
This gives $F_{cat} = 0.49$, $0.85$, and $0.983$, respectively.

The Wigner functions for the two-photon input state (Figs.~\ref{fig_n1_n2}(f,g,h)) are plotted for the copy
shift of $p^{(+)} = 4.08$. The parameters $\{y_m,\gamma\}$ are $\{3, 0.06\}$, $\{7, 0.14\}$, and
$\{15, 0.3\}$, with  $F_{cat} = 0.26$, $0.81$, and $0.955$, respectively.

In general, our results show that the non-Gaussian gate under consideration is able to effectively prepare
Scr{\"o}dinger cat states from arbitrary input state, which occupies a finite range in phase space.
As a ``perfect'' cat state, we imply here quantum superposition of undistorted copies of the input state
displaced by $\pm p^{(+)}$  along the momentum axis, where the superposition components acquire phase
factors $\exp(\pm i\theta)$, see~\eqref{theta}.

The criterion of how close is the gate output state to the ``perfect'' cat state is given by the fidelity $F_{cat}$~\eqref{cat_fidelity}. As it follows from Figs.~\ref{fig_n0}(d),~\ref{fig_n1_n2}(d),
and~\ref{fig_n1_n2}(h), the fidelity $F_{cat}$ decreases with the spread $2\Delta x$ of the input state
coordinate. This may be attributed to the increasing role of deformations due to higher degrees of $x$,
omitted in the Taylor expanded output state wave function~\eqref{cat_stphase} and~\eqref{factor_stphase}.
On the other hand, in order to improve the prepared cat state quality, one can choose the adequate gate
parameters from the fidelity $F_{cat}$ plots shown above.


 \section*{Conclusion}

We presented a detailed analysis of quantum statistical properties of the Schr{\"o}dinger cat states prepared by
the continuous-variable non-Gaussian gate. We have shown that the measurement-induced gate based on the
cubic phase state is able to produce conditionally the superpositions of two ``copies'' of an arbitrary input state
that occupies a finite area in phase space. The distance between the copies can be done as large as needed by
a proper choice of the gate parameters. The cat states quality is characterized by the fidelity between the gate
output state and the superposition of two symmetrically displaced undistorted copies of the input state. We found
criteria for the gate operation with high fidelity, and illustrated qualitative behavior of the output cat states in terms
of their Wigner functions for a wide range of parameters.

The gate exploits the same key elements, that is, the ancillary cubic phase state, the entangling $C_Z$ operation,
and homodyne measurement, as some other non-Gaussian CV schemes. A key feature of the regime where the cat-like
states arise is that the measurement provides multivalued information about the
target system physical variables. One can infer from our analysis that this feature may arise with other non-Gaussian
resource states, and with other types of measurements. This makes our approach potentially scalable and extendable
to more complex non-Gaussian networks.

The cat-breeding transformations of an arbitrary input state may be combined with standard Gaussian operations
such as displacement, rotation, squeezing, and shearing deformation. By establishing the measurement window
with a needed precision (which has  effect on the success probability), one can conditionally generate
complex cat-like structures in phase space. The multi-component Schr{\"o}dinger cat states are considered~\cite{Gottesman01,Mirrahimi14} as the logical qubit basis that ensures protection against errors of
different origin.

In general, we expect that a CV quantum network with embedded non-Gaussian gates of a general
kind may experience measurement-induced evolution to a Schr\"odinger cat state of an arbitrary complexity.

 \section*{Acknowledgments}

This research was supported by the Russian Foundation for Basic Research (RFBR) under the project 19-02-00204-a.

 \bibliography{}
 \bibliographystyle{plain}

 \end{document}